**Access models, authorship patterns, and citation impact in Ukrainian scholarly publishing (2020-2023)**

**Corresponding author:**
Myroslava Hladchenko
Center for R&D Monitoring, University of Antwerp, Belgium
hladchenkom@gmail.com, myroslava.hladchenko@uantwerp.be

**Abstract**
This study aimed to explore the relationship between access models, authorship patterns, and citation impact in Ukrainian research output from 2020 to 2023. The focus lied on scholars affiliated with the National Academy of Sciences of Ukraine (NASU) and universities. Findings highlight that OA articles constituted the majority of articles published by Ukrainian scholars between 2020 and 2023. This percentage reached 75.4% for NASU and 85.8% for universities. In both cases the rise occurred because of Gold OA and Hybrid Gold OA, the latter benefiting in part from Elsevier's waivers. Diamond OA prevailed for NASU and Gold OA for universities. The effects of Russia's full-scale invasion of Ukraine included (1) a decline in the share of articles in foreign journals for both NASU and universities, (2) a decrease in Gold OA in foreign journals and an increase in Gold OA in Ukrainian journals for universities, and (3) a rise in internationally co-authored Gold OA articles in foreign journals for both entities. Despite waivers for Gold OA provided by major publishers and an increase in Gold OA articles in Elsevier and Springer journals, MDPI and Aluna Publishing House remained the dominant publishers of Gold OA in foreign journals. Hybrid Gold OA, Bronze OA and Green OA articles in foreign journals had the highest citation impact. Citation impact of Gold OA outperformed Diamond OA. The study confirms the growing dominance of Gold OA, which suggests the need for sustainable OA models that ensure both equity and broad dissemination of research.

**Competing interests**
The author has no conflict of interests in the subject matter or materials discussed in this manuscript

**Introduction**
Academic publishing has been increasingly shifting toward open access (OA), which guarantees the free availability of research outputs (BOAI, 2012; Solomon & Björk, 2012; Kaliuzhna et al., 2025). Research can become freely accessible in two main ways: through self-archiving (Green OA) or by journals providing open access directly (Tennant et al., 2016).

Depending on publishing conditions, OA journals can be divided into Diamond OA, which do not charge either authors or readers, and Gold OA, which do. Diamond OA goes beyond simply removing APCs: it supports equity, treats knowledge as a public good, and encourages diversity and multilevel cooperation in scholarly publishing (Simard et al., 2024). Hybrid Gold OA journals may also provide temporary or permanent free access without a common license, often referred to as Bronze OA. Importantly, journals charging APCs currently account for the majority of OA publications (Crawford, 2018).

However, Gold and Hybrid Gold OA models require substantial funding. The APC system not only creates inequalities in publishing (Brainard, 2024; Demeter & Istratii, 2020; Olejniczak & Wilson, 2020; Siler et al., 2018), but also raises concerns about whether allocating research budgets to APCs represents a sustainable use of funding. Moreover, the APC model—while intended to make research freely available—can be exploited by publishers prioritising profit over rigorous peer review (Bohannon, 2013). Despite these challenges, countries worldwide are increasingly embracing OA in national strategies and policies.

This study examines Ukraine's experience with OA publishing. Following the dissolution of the Soviet Union in 1991, Ukrainian publication requirements emphasised articles in domestic journals. Most Ukrainian journals charged APCs, while making their content publicly accessible on journal websites (Hladchenko & Moed, 2021). As a result, OA publishing quickly has been well established in Ukraine, and paying APCs has turned into an institutionalised practice within Ukrainian scholarly communication. Since 2010, the Ministry of Education and Science began monitoring the global rankings of Ukrainian universities. Because these rankings assess research performance largely based on articles published in Scopus- and WoS-indexed journals, the Ministry introduced new policies in 2013 requiring such publications for doctoral degrees (Ministry of Higher Education and Science, Youth and Sport of Ukraine, 2012; Hladchenko & Westerheijden, 2019; Hladchenko, 2023; 2024). These requirements were later extended to academic promotion, including appointments to associate professor and professor positions (Ministry of Education and Science of Ukraine, 2016; Abramo et al., 2023). This policy shift was intended to increase the international visibility and ranking positions of Ukrainian universities.

In 2013, Ukrainians launched the Revolution of Dignity after the pro-Russian president refused to sign the EU Association Agreement. Although the revolution succeeded, Russia responded by invading Ukraine in 2014. In 2022, the invasion escalated into a full-scale war across the entire territory of Ukraine. This war has severely disrupted academic life (Oleksiyenko et al., 2021; 2023; Fiialka, 2022; Kiselyova & Ivashchenko, 2024; De Rassenfosse et al., 2023). Everyday survival has taken priority over research, as scholars face bombing, blackouts, and the consequences of economic collapse caused by the diversion of national resources to defense (BBC, 2024). As of 2024, 1,518 Ukrainian scholars have joined the army to defend Ukraine (Cabinet of Ministers of Ukraine, 2024). A total of 146 Ukrainian scholars were killed, comprising individuals who served in the army as well as those killed in Russian missile attacks (Novynarnia, 2024).

Despite the war, Ukrainian scholars continue to publish, which deserves recognition. At the same time, the Ukrainian government not only defends against Russia's invasion but also advances policies aimed at EU integration. Notably, in 2022 the Cabinet of Ministers of Ukraine approved a National Plan on Open Science (Cabinet of Ministers of Ukraine, 2022). This policy was adopted as part of Ukraine's agreement to participate in the Horizon Europe and Euratom Framework Programmes (Ministry of Education and Science of Ukraine, 2022).

This study has two main objectives: first, to explore the relationship between access models, authorship patterns, and citation impact; and second, to examine how Russia's full-scale war against Ukraine affected open-access scholarly publishing by Ukrainian scholars. The analysis focuses on scholars affiliated with either the National Academy of Sciences of Ukraine (NASU) or universities, as these two sectors employ the majority of Ukrainian researchers. Data were drawn from the CWTS in-house Scopus database and include articles published before (2020–2021) and after (2022–2023) the start of Russia's full-scale invasion.

## Publication access models

In 2015, open access (OA) publications constituted 45% of scholarly output registered in Crossref (Piwowar et al., 2018). Publications can become freely accessible online in two main ways. First, authors can self-archive their work as preprints on the web, in institutional repositories, or on personal websites (Tennant et al., 2016). Self-archiving rights depend on journal or publisher policies (Laakso, 2014). Some publishers impose embargo periods before articles can be deposited in public repositories, typically to protect subscription revenues. Notably, Elsevier and Springer do not require embargoes for posting preprints. Björk et al. (2014) found that 62% of journals from the top 100 publishers indexed in Scopus allow immediate Green OA self-archiving, 4% impose a six-month embargo, and 13% require a 12-month embargo.

Second, journals themselves may provide OA to publications. Scopus distinguishes between Bronze OA, Gold OA, Green OA, and Hybrid Gold OA. Gold OA journals include both those that charge Article Processing Charges (APCs) and those that do not, with the latter referred to as Diamond OA. The academic community increasingly views Diamond OA as a potential solution to inequities created by APC-based models (Simard et al., 2024). Diamond OA journals are particularly common in the social sciences and humanities (Taubert et al., 2024). They are typically owned by governmental or academic institutions, with financial stability being a key factor for their sustainability (Yoon et al., 2024).

Journals that charge APCs constitute 30% of OA journals but publish 56% of research output (Crawford, 2018). APCs are charged mandatorily by Gold OA journals and optionally by Hybrid Gold OA journals. Subscription-based journals have not abandoned their traditional model but often offer an OA option, creating a hybrid model (Zhang, 2024; Laakso & Björk, 2016). Hybrid Gold OA allows articles in subscription journals to be made freely accessible for a fee. These journals often charge higher APCs because they are older, more prestigious, and have higher impact factors than Gold OA journals (Pinfield et al., 2017; Björk & Solomon, 2017). For instance, in medicine, hybrid journals charge higher average APCs than Gold OA journals (Siler et al., 2018). Hybrid journals are generally more reputable and well-established, attracting more interest from both authors and readers (Brainard, 2024). However, the Hybrid Gold OA model has been criticised as exploitative.

The APC model creates inequalities in knowledge production, preventing a large segment of scholars—especially those in developing countries or affiliated with less wealthy institutions—from publishing (Demeter & Istratii, 2020; Olejniczak & Wilson, 2020). This issue is particularly acute in medicine, where APCs are the highest (Siler et al., 2018; Asai, 2019). APCs impose a significant burden on national science systems and institutions, prompting the search for alternative publishing models (Zhang et al., 2022; Pinfield et al., 2016). Nation-states and research institutions initially considered OA publishing as an alternative to subscriptions, aiming both to make research output publicly available and to reduce subscription costs. However, for publishers, APC-based OA proved more profitable than subscriptions, thereby increasing the costs for institutions (Maddi, 2020; Maddi & Sapinho, 2021). This has negative consequences for science and scholars (Haustein et al., 2024; Alonso-Álvarez et al., 2024). Another concern is the potential deterioration of publication quality, as high-impact journals may become more lenient and accept lower-quality articles under an APC-based OA model than they would under subscription-based access (Van Vlokhoven, 2019).

Open access was intended to address accessibility, affordability, and equity, but it has failed to fully solve the latter two problems. OA advocates did not maintain ownership of the movement, allowing publishers to repurpose it for profit. Without addressing affordability, the equity problem cannot be solved (Anderson, 2023).

To reduce financial burdens on researchers and institutions while ensuring free public access, institutions and countries increasingly encourage scholars to post preprints (ROARMAP database), which constitutes Green OA. Preprints enable early discovery, fast and wide dissemination, free access, and enhanced visibility of publications (Chiarelli et al., 2019; Wang et al., 2020), which can increase citation counts (Wang et al., 2018; Piwowar et al., 2018; Colavizza et al., 2024).

When choosing a journal, authors prioritise fit, quality, and speed of publication, with OA being a less important but still significant factor (Solomon & Björk, 2012). Despite the financial burdens, scholars in medicine affiliated with prestigious universities often prefer Gold OA over Diamond OA (Siler et al., 2018).

Prior studies highlighted an OA citation advantage (McCabe & Snyder, 2014; Archambault et al.,2014; Ottaviani, 2016). Piwowar et al. (2018) revealed that OA articles receive 18% more citations than average while closed ones are cited 10% below the world average. However, the citation impact varies across the types of OA. Green OA and Hybrid OA are cited 33% and 31% above the world average. Gold OA articles are cited 17% below the world average and 9% below the closed articles. Bronze OA articles are cited 22% above the world average. Dorta-González & Dorta-González, (2022) revealed that Hybrid Gold OA articles are cited on average twice as Gold OA ones. However, even if the articles are not published in OA source, the use of open-access repositories considerably increases the citations received, especially for those articles without funding. They are cited on average twice as closed ones (Dorta-González & Dorta-González, 2022) The comparison of Gold OA and Diamond OA journals in engineering revealed that the former have higher averages in citations, although they show more significant variability (Pilatti et al., 2024).

Dorta-González & Dorta-González, (2022) state that OA publications are cited more than closed ones not only because of free access to them but also because these publications resulted from well-funded studies. Nowadays research funders stipulate that, unless a researcher has published in an OA journal, a green OA copy must be made available (Björk et al., 2014; Else, 2018; van Noorden, 2018). For example, since 2018, Spain has obliged publications resulting from projects funded either by public or private agencies to be in open access (Plan, 2023). Funded publications tend to obtain more citations than non-funded ones (Wang & Shapira, 2015; Sandstrom, 2009). That is why as the articles from funded studies tend to be published OA, they are more likely to receive more citations than closed publications. Comparisons between Gold OA and Diamond OA journals in engineering indicate that Gold OA journals achieve higher average citations, although they exhibit greater variability (Pilatti et al., 2024).

**Open Science in Ukraine**

After the fall of the Soviet Union in 1991, Ukraine maintained a clear division between primarily teaching-oriented higher education institutions and the research institutes of the National Academy of Sciences of Ukraine (NASU). This institutional differentiation contrasted with Western countries, where, even in systems with non-university research institutes (e.g., Germany), universities typically combined both teaching and research functions (Clark, 1983).

NASU inherited a hierarchical governance structure from the Soviet period. The Board of Academicians (Presidium) manages the academy, which comprises research institutes organized into fourteen disciplinary sections: mathematics; information technology; mechanics; physics and astronomy; earth sciences; material sciences; energy; nuclear physics and energy; chemistry; biology, physiology, and molecular biology; general biology; economics; history, philosophy, and law; literature, language, and art. Physical and technical sciences dominate in terms of the number of researchers and research institutes (Josephon & Egorov, 1997). Regarding higher education institutions, most were renamed as universities in the early 1990s. Legislation authorised and encouraged these universities to engage in both fundamental and applied research. In 2014, the maximum annual teaching hours were reduced from 900 to 600 to allow academics more time for research (Parliament of Ukraine, 2014; Hladchenko & Westerheijden, 2021).

After 1991, publication requirements for doctoral degrees and promotion to professor initially focused solely on articles published in Ukrainian journals. In 2013, the Ministry of Education and Science introduced the requirement to publish in Ukrainian journals indexed in Scopus or WoS for doctoral degrees (Ministry of Education and Science, Youth and Sport of Ukraine, 2012). Subsequently, these publications were included in licensing requirements for study programs (Cabinet of Ministers of Ukraine, 2015), criteria for academic titles such as associate professor and professor (Ministry of Education and Science of Ukraine, 2016), and the research assessment of higher education institutions (Cabinet of Ministers of Ukraine, 2017). Recent studies indicate that the number of articles published by Ukrainian scholars in Scopus-indexed journals has increased. However, Ukrainian academics published extensively in local Scopus-indexed journals (Nazarovets, 2020; 2022).

In recent decades, Ukraine has implemented several initiatives to promote Open Science. In 2007, the Parliament of Ukraine approved the law *On Main Principles of the Development of Information Society in Ukraine in 2007–2015*, which required the establishment of technical infrastructure and electronic information resources in archives, libraries, museums, and research organizations to store research results and provide open access to findings from state-funded research. In 2016, the Cabinet of Ministers of Ukraine approved the establishment of the National Repository of Academic Texts (NRAT) (Cabinet of Ministers of Ukraine, 2016), which houses doctoral theses, scientific publications, and data on research projects funded by the Ministry of Education and Science of Ukraine (Ministry of Education and Science of Ukraine, 2022a). Through NRAT, Ukraine can integrate its research output into the European Open Science Cloud (EOSC) (Chmyr, 2019). Although Ukrainian legislation does not mandate universities to maintain repositories, many institutions have done so voluntarily.

In 2022, the Cabinet of Ministers of Ukraine approved a national plan on open science (Cabinet of Ministers of Ukraine, 2022). This policy was adopted under Ukraine's participation in the Framework Programme Horizon Europe and Euroatom, which, in Article 6, requires support for open science practices (Ministry of Education and Science of Ukraine, 2022). Ukrainian academic journals have a long tradition of publishing with open access (Hladchenko & Moed, 2022). When journals charge an APC, it is typically covered by academics from their own remuneration. Analysing Ukrainian research output from an access perspective, Kaliuzhna and Hauschke (2024) found that 71.5% of publications by Ukrainian scholars in 2012–2021 were in open-access sources, with 60.9% of these in Ukrainian journals. Hladchenko (2025) noted that in 2018–2022, Ukrainian scholars increasingly published in foreign Gold OA journals.

Russia's full-scale invasion of Ukraine also prompted responses from the international community, including international publishers (Nazarovets & Teixeira da Silva, 2022). In particular, international publishers introduced waivers for Gold OA journals for corresponding authors affiliated with Ukrainian institutions. These publishers include Wiley, IOP Publishing, Sage, Springer, and Taylor & Francis. In February 2022, Elsevier abolished APCs for both Gold OA and Hybrid Gold OA; however, since October 2023, this option is no longer available for hybrid journals (Elsevier, 2025; Ministry of Education and Science of Ukraine, 2022a). Oxford University Press and Frontiers review waivers on a case-by-case basis. Walter de Gruyter GmbH provides a 50% discount for Ukraine as a lower-middle-income country. PLOS ONE allows corresponding authors affiliated with Ukrainian institutions to publish without a fee. The Polish ALUNA Publishing House offers a partial discount, reducing APCs from €600 to €350. MDPI has not announced formal waivers for Ukrainian scholars but provides reviewer vouchers that partially or fully cover APCs, and guest editors of special issues can publish free of charge.

**Data and methods**

First, articles published between 2020 and 2023 with at least one author (or co-author) affiliated with Ukrainian institutions were extracted from the CWTS in-house Scopus database. This database was selected for the initial dataset because it provides structured affiliation data. Based on Ukrainian affiliations, all articles (N = 61,105) were divided into three categories: authored by scholars affiliated with the NASU (N = 17,554), authored by scholars affiliated with universities (N = 45,496), and authored by scholars affiliated with other institutions (N = 4,098). Articles co-authored by scholars affiliated with both NASU and universities were counted twice.

Second, the EIDs of the articles were exported to Scopus to obtain information on their access model. Scopus defines the following types of OA: Gold Open Access, Hybrid Gold Open Access, Green Open Access, and Bronze Open Access. However, Scopus does not distinguish between Gold OA journals that charge APCs and those that do not. Gold OA journals without APCs are referred to as Diamond OA. To differentiate between Gold OA (charging APCs) and Diamond OA (not charging APCs), data from the Directory of Open Access Journals (DOAJ) were used, and journals not listed in DOAJ were manually checked. Journals providing open access without explicitly charging APCs were classified as Diamond OA. Journals operating under a "Subscribe to Open" (S2O) model without APCs, such as the *Journal of Fractal Geometry*, were also classified as Diamond OA.

Scopus defines Bronze OA as: "Published version of record or manuscript accepted for publication, for which the publisher has chosen to provide temporary or permanent free access, with no Creative Commons license attached." A check of publications marked as Bronze OA revealed that some belonged to Diamond OA, Gold OA, or Hybrid Gold OA journals. Accordingly, these were recategorized, while the Bronze OA designation was retained for publications in Hybrid Gold OA journals. A manual random check of 3,002 journals classified by Scopus as closed-access revealed that 2,467 were indeed closed (subscription-based or hybrid), while the remaining were Diamond OA or Gold OA. Similarly, publications marked as Green OA but belonging to Diamond or Gold OA journals were reclassified according to the journal's access type. Notably, 98.8% of Green OA articles (1,934 out of 1,957) published by NASU and university scholars were in hybrid journals, while the remainder were in closed journals.

Scopus defines Bronze OA as: "Published version of record or manuscript accepted for publication, for which the publisher has chosen to provide temporary or permanent free access, with no Creative Commons license attached." A check of publications marked as Bronze OA revealed that some belonged to Diamond OA, Gold OA, or Hybrid Gold OA journals. Accordingly, these were recategorized, while the Bronze OA designation was retained for publications in Hybrid Gold OA journals. A manual random check of 3,002 journals classified by Scopus as closed-access revealed that 2,467 were indeed closed (subscription-based or hybrid), while the remaining were Diamond OA or Gold OA. Similarly, publications marked as Green OA but belonging to Diamond or Gold OA journals were reclassified according to the journal's access type. Notably, 98.8% of Green OA articles (1,934 out of 1,957) published by NASU and university scholars were in hybrid journals, while the remainder were in closed journals.

Authorship patterns were classified into two categories: Ukrainian (UKR), authored solely by scholars affiliated with Ukrainian institutions, and internationally co-authored (INTER), which included at least one foreign institutional affiliation. To identify Ukrainian journals, the list compiled by Hladchenko and Moed (2021) was used, supplemented with data from the DOAJ and manual verification. Journals published by Springer but managed by Ukrainian editorial boards or institutions were also considered Ukrainian for this study. For example, the *Journal of Mathematical Sciences* includes Series A and B, with Series B comprising English-language translations from 12 Russian- and Ukrainian-language source journals.

To measure citation impact, the field-normalised citation impact (FNCI) was calculated for each article. Citations were normalized by discipline, publication year, and document type. Specifically, the number of citations to each article was divided by the world average number of citations for articles of the same discipline, year, and type. Articles classified in multiple disciplines were fractionalised accordingly.

**Results**
**Distribution of articles published by Ukrainian scholars across access models**
Figure 1 presents the distribution of articles published by Ukrainian scholars between 2020 and 2023 across different access models. Open Access (OA) articles constituted the majority throughout the period, with their share increasing from 78.0% to 83.8%. This growth was mainly driven by a rise in Hybrid Gold OA, accompanied by a decline in closed articles in hybrid journals and a slight increase in Gold OA. Among the access models, Gold OA accounted for the largest share, rising from 41.6% to 44.5%. Diamond OA ranked second, remaining relatively stable at around 28.5%.

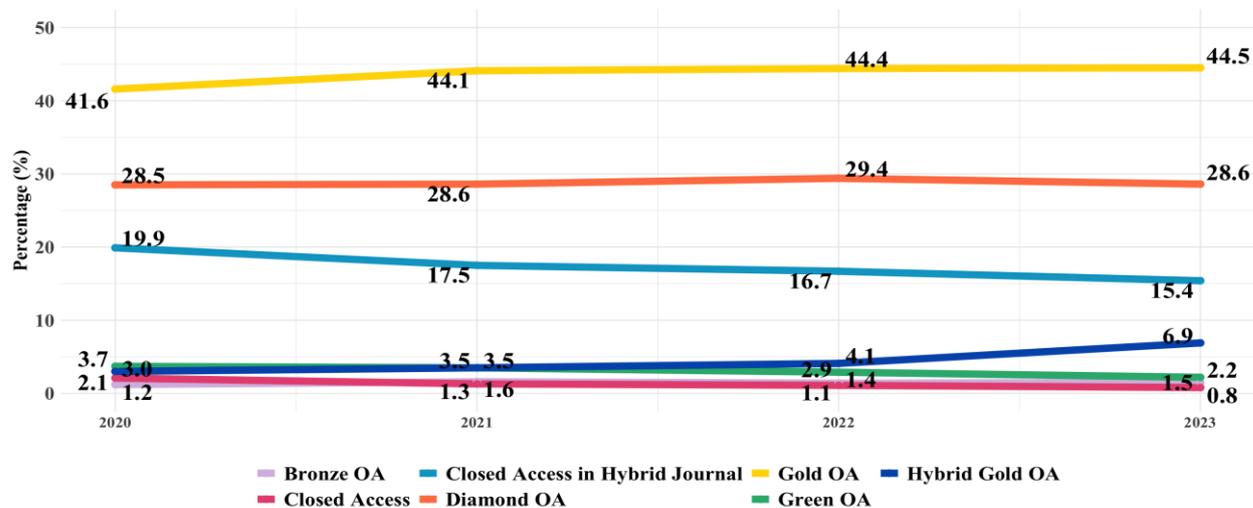

Fig. 1 Distribution of articles published by Ukrainian scholars across access models

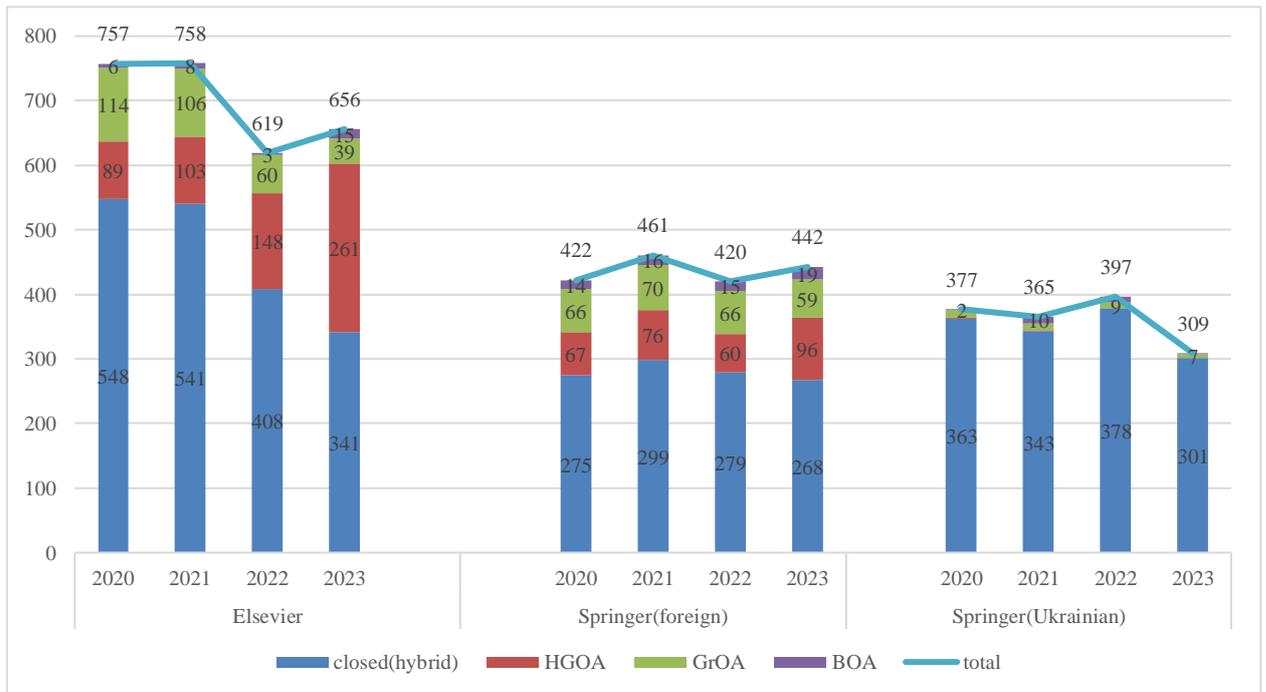

Fig. 2 Articles in hybrid journals of Springer and Elsevier

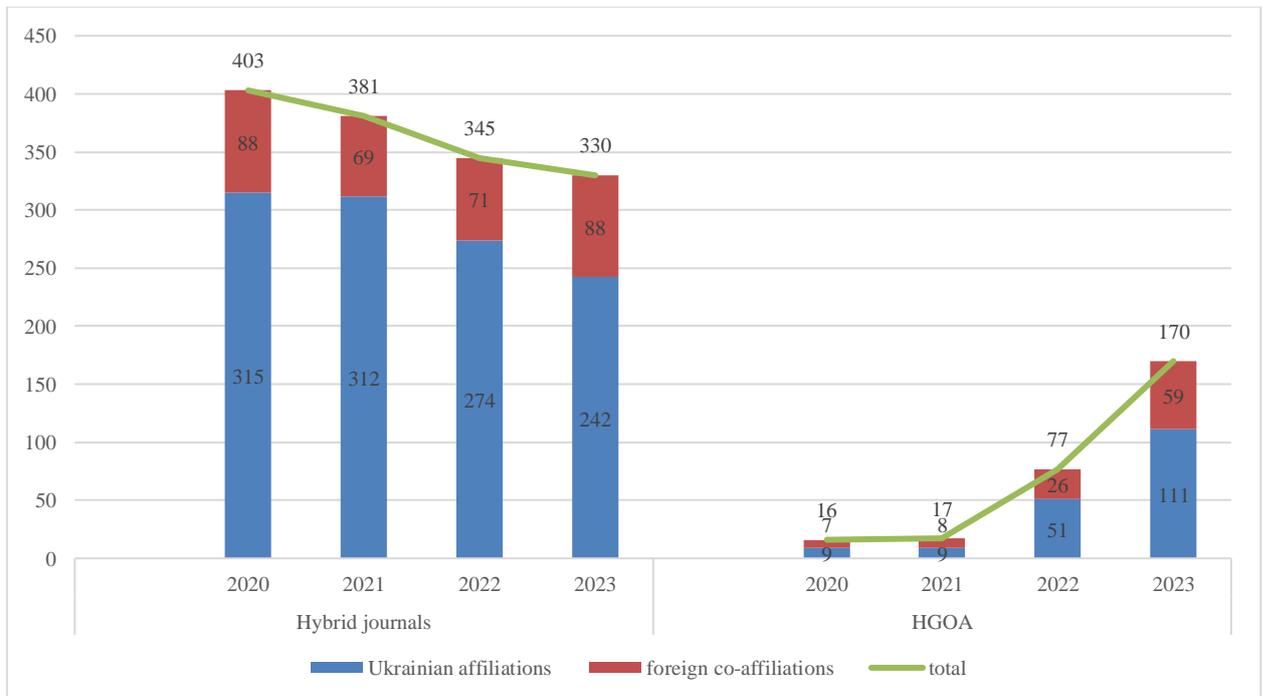

Fig. 3 Corresponding authors affiliated with Ukrainian institutions in hybrid journals of Elsevier (the number of articles is 2663, 95,4% of the total)

As the proportion of Hybrid Gold OA articles increased, Figure 2 shows the number of articles published in hybrid journals by major publishers, including Elsevier and Springer (with Nature), which represent the leading publishers of this journal type. The figure indicates that, although the number of Hybrid Gold OA articles in Elsevier journals increased, the total number of articles declined following the start of Russia's full-scale invasion.

Figure 3 shows that between 2020 and 2023, the total number of articles with corresponding authors affiliated with Ukrainian institutions in Elsevier's hybrid journals fell from 403 to 330. Meanwhile, the share of articles with corresponding authors holding foreign co-affiliations increased. Notably, the number of Hybrid Gold OA articles with Ukrainian-affiliated corresponding authors rose sharply from 16 to 170. However, this figure remained substantially lower than the total number of articles with Ukrainian corresponding authors, suggesting both the limited duration of Elsevier's APC waiver and that not all Ukrainian-affiliated corresponding authors took advantage of this opportunity.

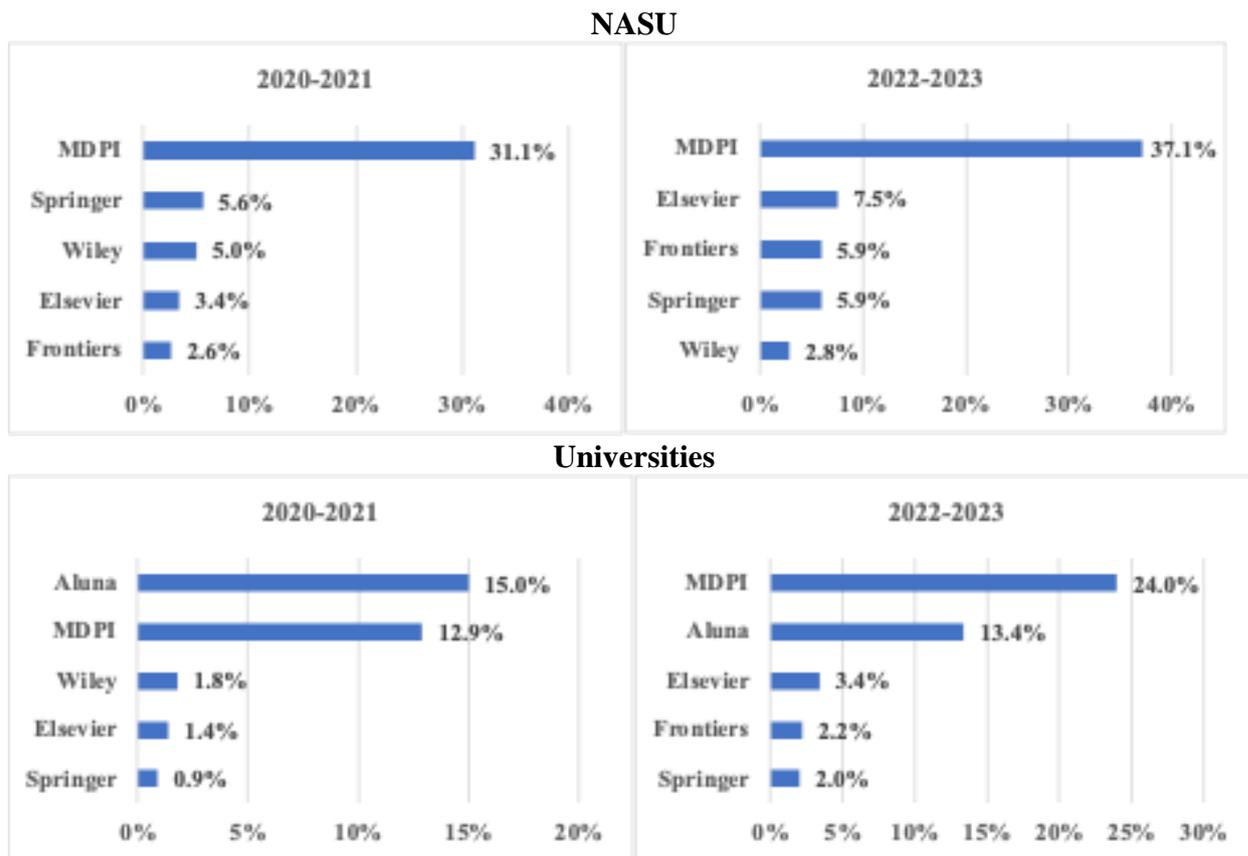

Fig. 4 Top five Gold OA publishers in foreign journals (share of total articles published in Gold OA in foreign journals)

Figure 4 shows that MDPI was the leading Gold OA publisher for NASU in both periods, with its share increasing from 31.1% to 37.1%. The share of Springer articles remained almost unchanged, while those of Elsevier and Frontiers journals rose slightly. For universities, Aluna Publishing House, which specialises in medical journals, ranked first in 2020–2021, followed by MDPI. In 2022–2023, however, MDPI took the lead with a significant increase in its share, while Aluna's declined. The shares of Springer, Elsevier, and Frontiers journals also rose modestly during this period.

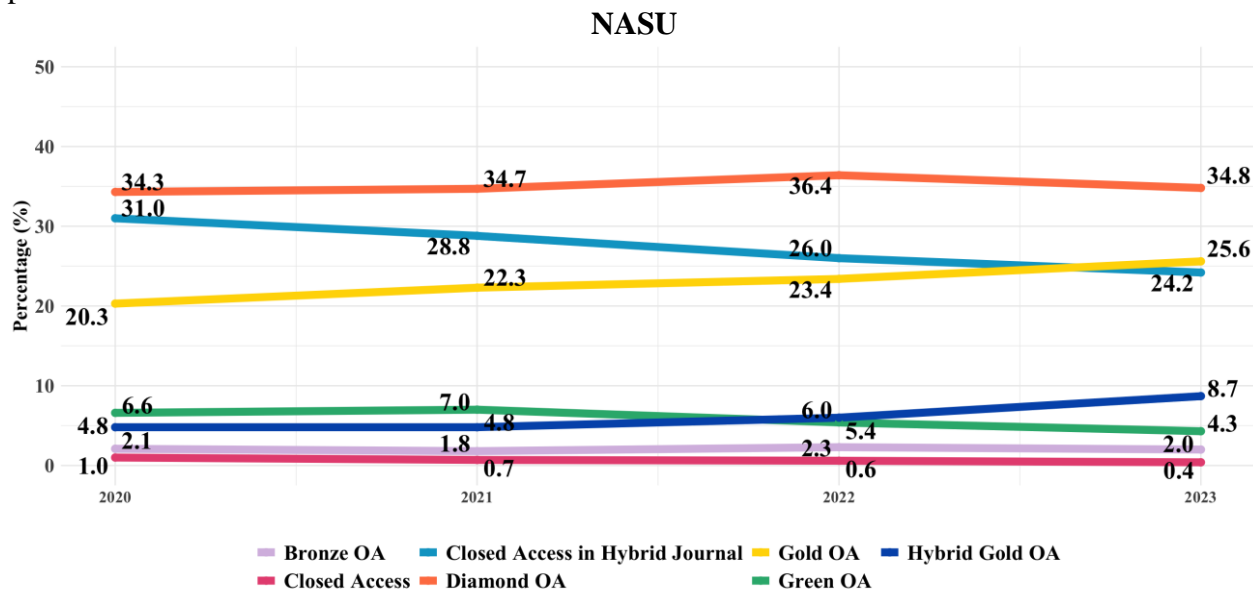

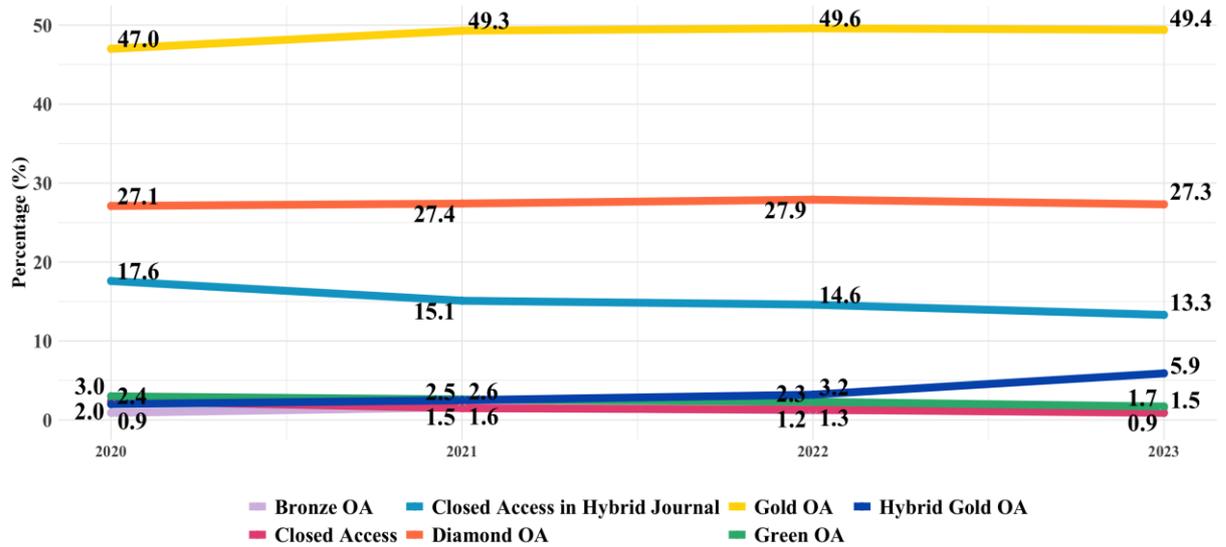

Fig. 5 Distribution of articles across access models

Figure 5 illustrates the distribution of articles across access models. For NASU, the percentage of OA articles increased from 68.0% to 75.4%. The increase occurred due to the rise in the percentage of Gold and Hybrid Gold OA. This implied a simultaneous decrease in closed articles published in hybrid journals. The largest share of OA articles constituted those published in Diamond OA, which remained relatively stable, fluctuating at 34.0%.

For universities, the share of OA articles also increased, rising from 80.0% to 85.8%. The growth was largely due to an increase in Hybrid Gold OA, from 2.4% to 5.9%, accompanied by a decrease in closed articles in hybrid journals, from 17.6% to 13.3%. In contrast to NASU, where Diamond OA dominated, Gold OA was the most prevalent model among universities, maintaining a stable share of approximately 50.0%. Meanwhile, Diamond OA fluctuated around 27.0%.

**Distribution of articles across access models in Ukrainian and foreign journals**

Figure 6 highlights that the majority of articles authored by scholars affiliated with NASU were published in foreign journals, with the share decreasing from 56.5% to 54.9%. In both periods, closed articles in hybrid journals remained predominant, though their share declined, while Hybrid Gold OA increased. Diamond OA articles represented the largest share in Ukrainian journals.

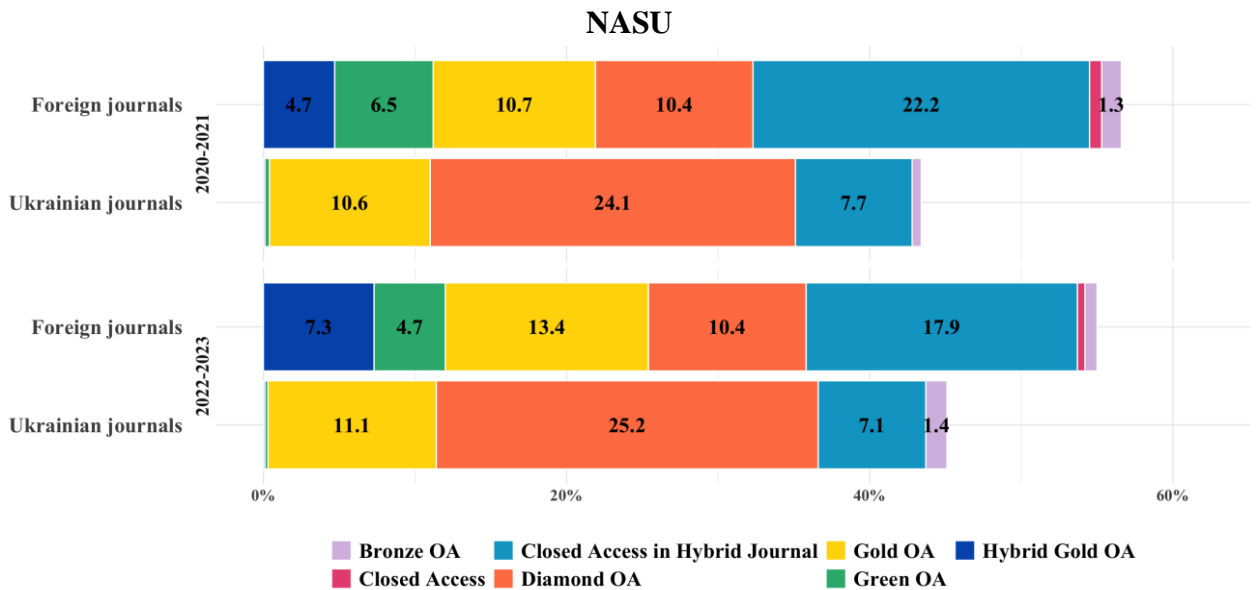

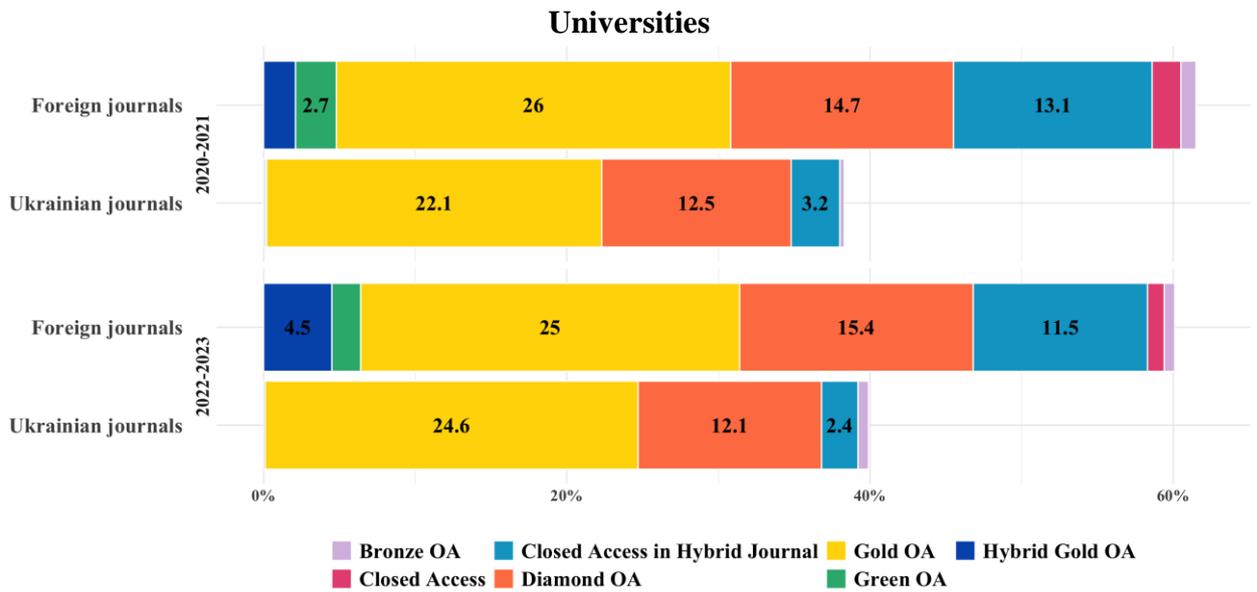

Fig. 6 Distribution of articles across access models in Ukrainian and foreign journals

For universities, the share of articles in foreign journals declined from 61.6% to 60%. Within foreign journals, Gold OA articles constituted the largest portion, declining from 26.0% to 25.0%. The share of closed articles in hybrid journals decreased, while Hybrid Gold OA increased from 2.1% to 4.5%. Diamond OA rose only slightly, from 14.5% to 15.3%.

In Ukrainian journals, Gold OA also prevailed and increased from 22.1% to 24.6%, while the shares of other access models remained largely stable.

**Access models and authorship patterns in foreign journals**

Articles authored solely by Ukrainian scholars accounted for the majority of publications in Ukrainian journals in both periods. For NASU, they represented 87.8% of articles in 2020–2021 and 85.9% in 2022–2023, while for universities the shares were 89.3% and 87.5%, respectively. Because internationally co-authored articles made up only a small proportion in Ukrainian journals, this subsection focuses on publications in foreign journals.

Figure 7 shows authorship patterns in NASU publications across different access models in foreign journals. International collaborations dominated in all categories except closed access. In 2022–2023, the shares of Gold OA and Hybrid Gold OA increased, reflecting growth in both domestically authored and internationally co-authored articles.

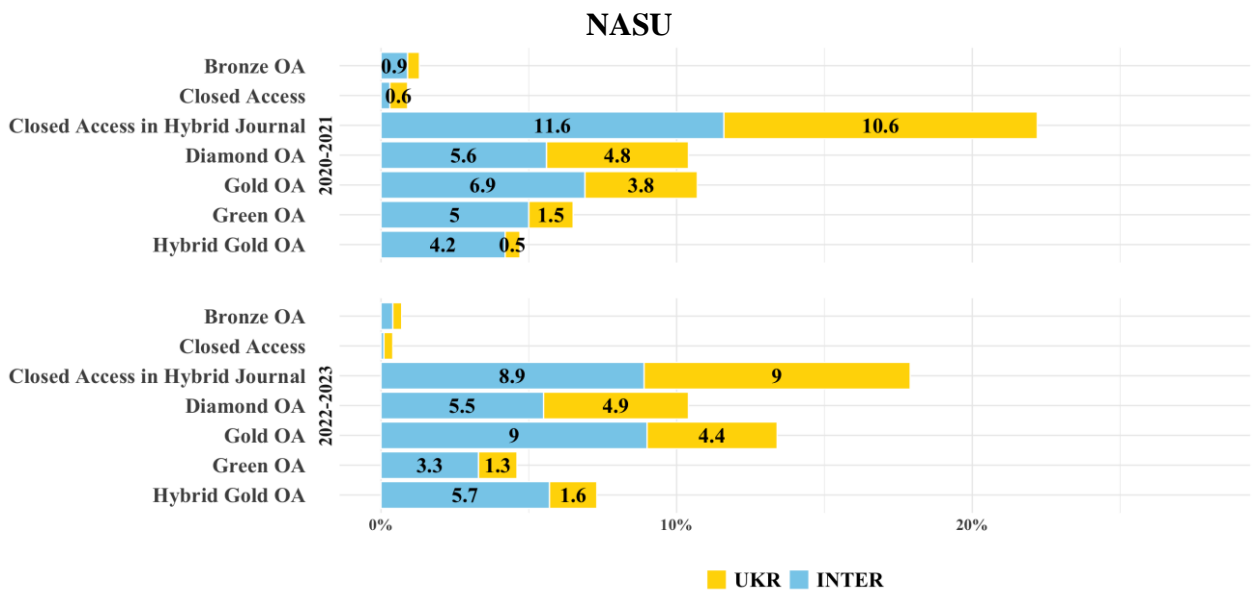

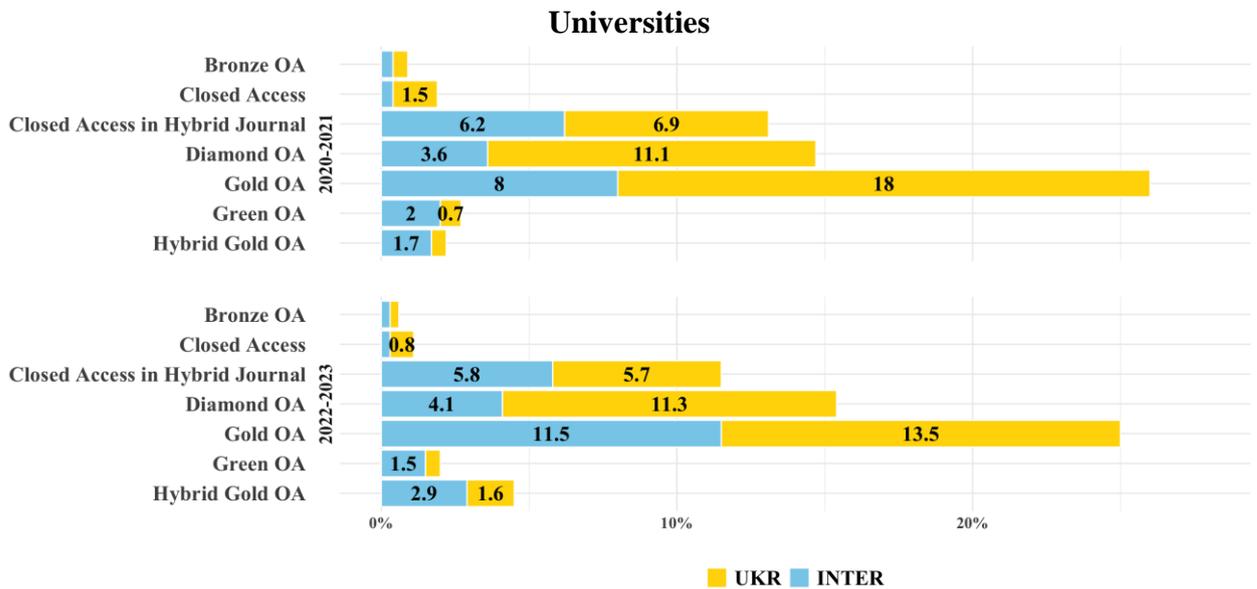

**Fig. 7** Distribution of authorship patterns across access models in foreign journals

In contrast to NASU, for universities, articles authored solely by Ukrainian scholars dominated Diamond OA, Gold OA, and closed-access publications (Figure 10). This dominance in closed-access articles in hybrid journals disappeared in 2022–2023. Internationally co-authored articles prevailed in Hybrid Gold OA and Green OA in both periods. Gold OA remained the largest category overall. The share of internationally co-authored Gold OA articles increased from 8.0% to 11.5%, while the proportion of articles authored solely by Ukrainian scholars declined from 18.0% to 13.5%. This decrease may reflect the economic difficulties faced by Ukrainian scholars due to the war, which affected their ability to cover APCs. Similar to NASU, the share of Hybrid Gold OA articles rose slightly in both authorship categories in 2022–2023.

**Relationship between access models and citation impact**

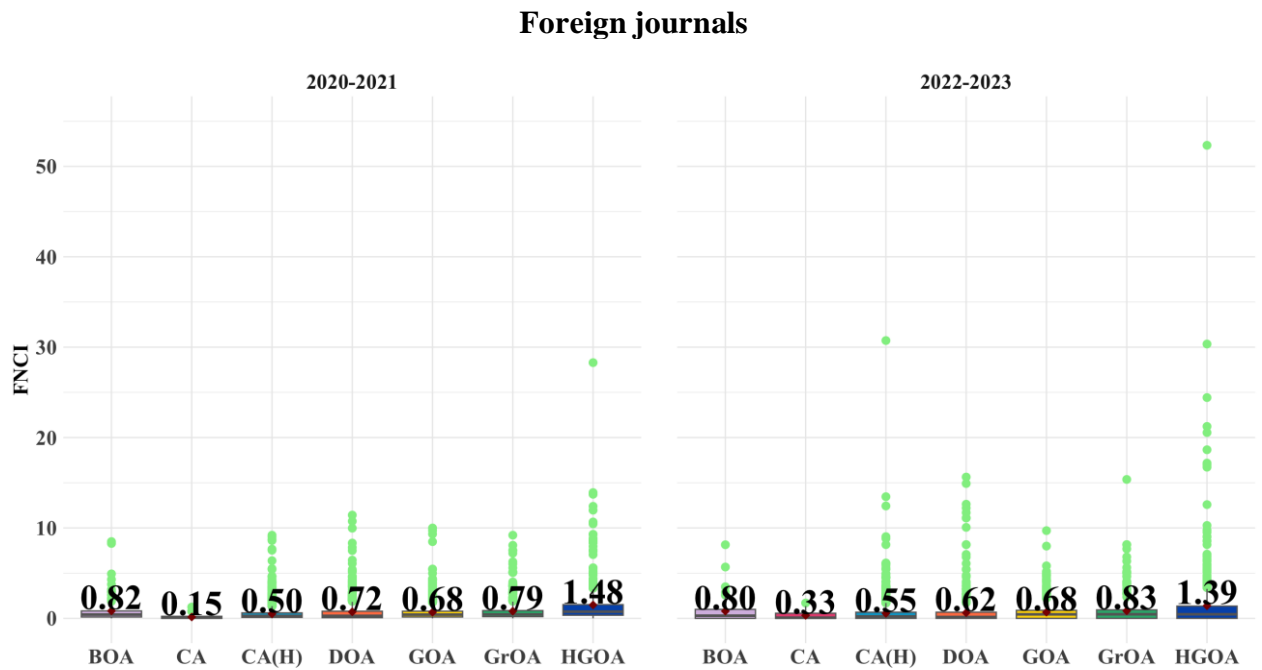

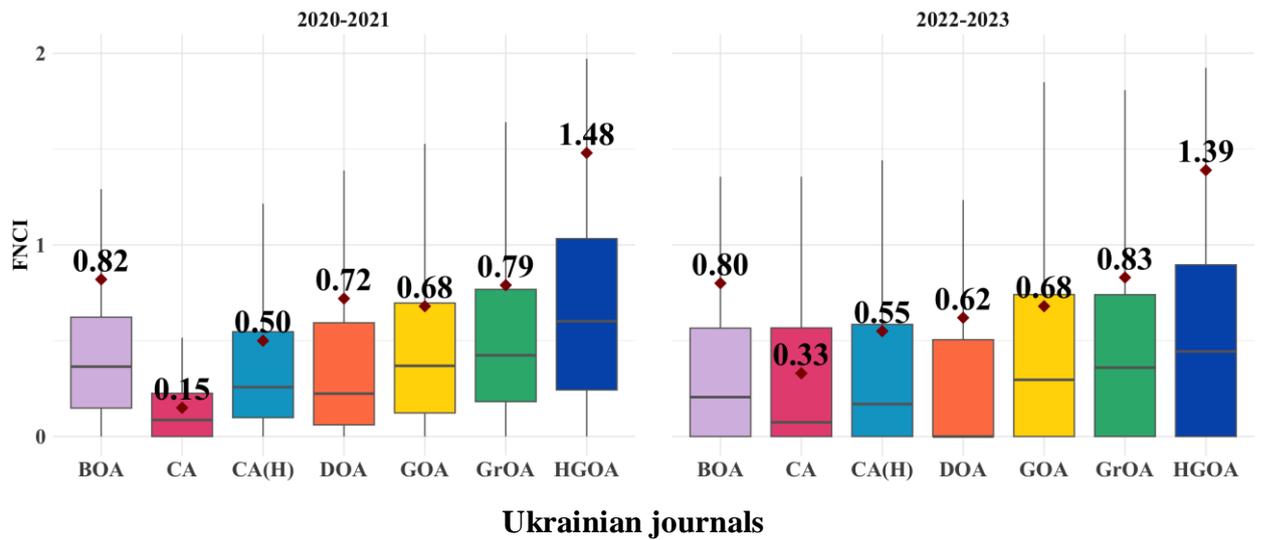

**Ukrainian journals**

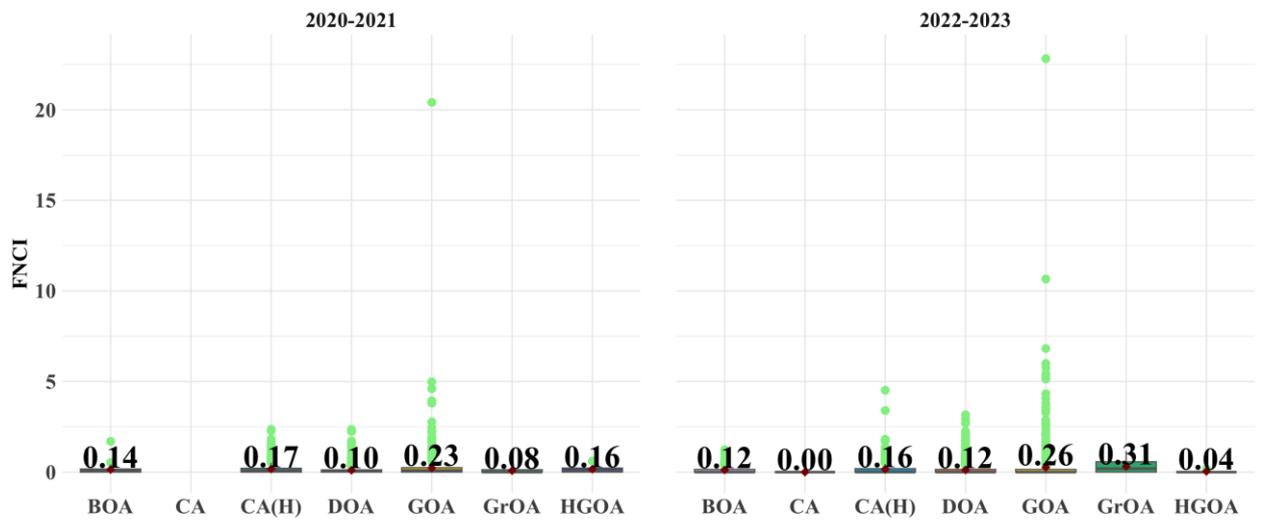

Fig. 8 Citation impact across access models in foreign and Ukrainian journals (NASU)

Figure 8 illustrates the citation impact of articles authored by NASU-affiliated scholars across access models in foreign and Ukrainian journals. In both periods, Hybrid Gold OA exhibited the highest average FNCI, followed by Bronze OA and Green OA. However, these high averages were driven not by the majority of articles but by a few highly cited outliers. In 2020–2021, Diamond OA articles had a higher average FNCI than Gold OA, but this trend reversed in 2022–2023. In Ukrainian journals, average FNCI values were substantially lower across all access models compared to foreign journals.

**Foreign journals**

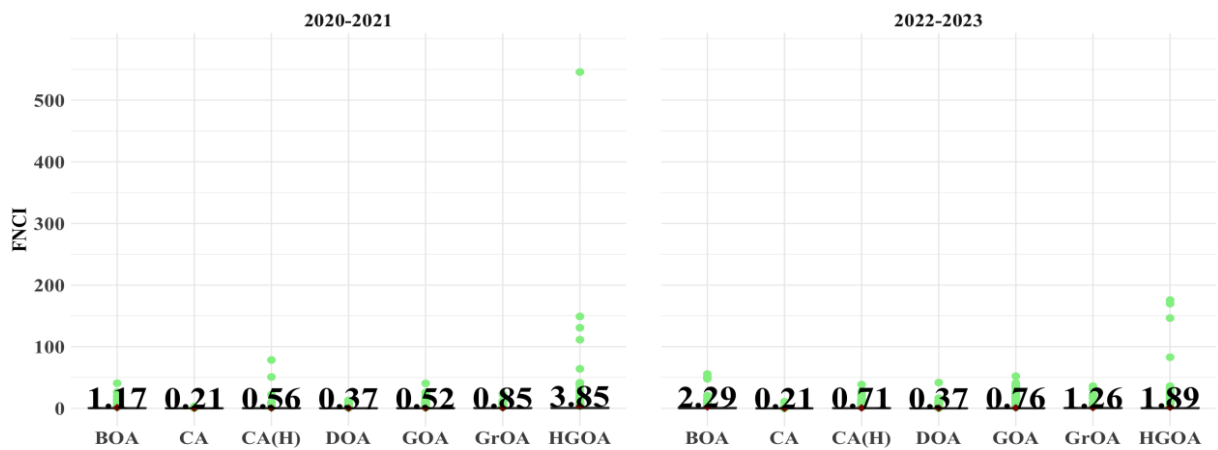

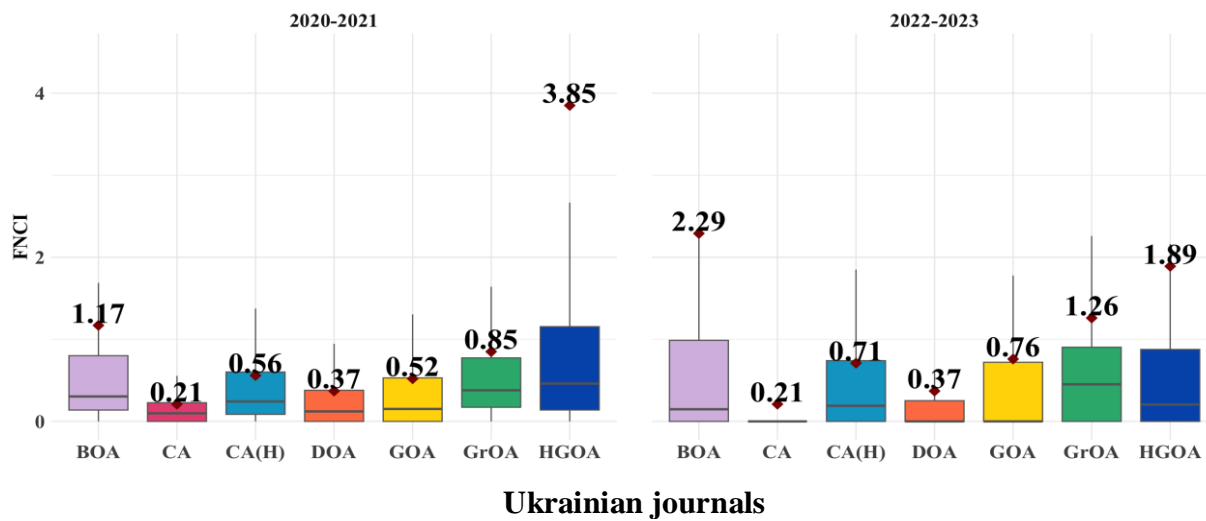

**Ukrainian journals**

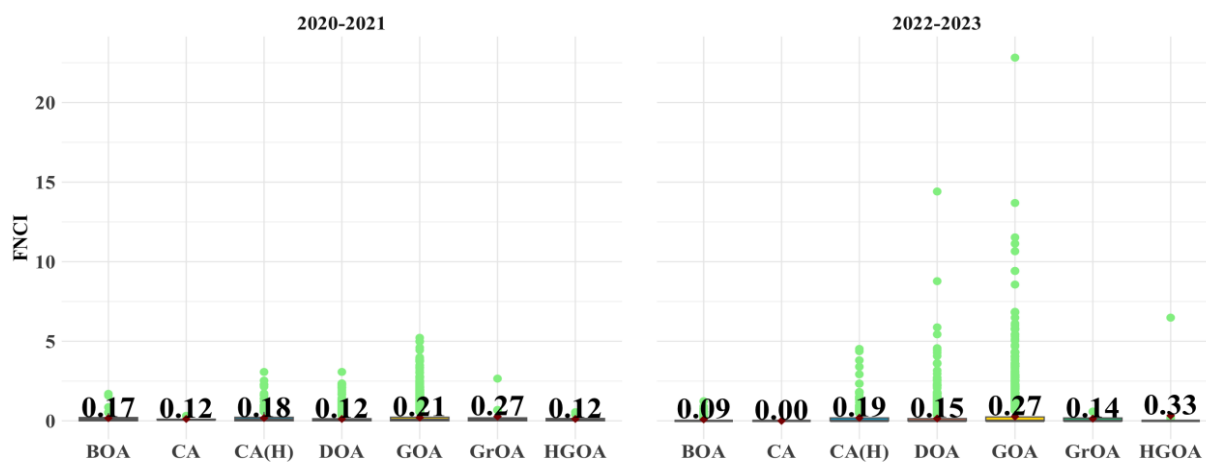

Fig. 9 Citation impact across access models in foreign and Ukrainian journals (universities)

Figure 9 illustrates the citation impact of articles authored by university-affiliated scholars across access models in foreign and Ukrainian journals. Similar to the NASU, Hybrid Gold OA, Bronze OA, and Green OA exhibited the highest average FNCI values. These elevated averages were largely driven by highly cited outliers rather than the majority of articles. Diamond OA consistently had a lower average FNCI than Gold OA in both periods, with the gap widening in 2022–2023. Closed-access articles had the lowest average FNCI for both NASU and universities. However, while for NASU the average FNCI of closed-access articles in hybrid journals was lower than that of Diamond OA and Gold OA, for universities it was higher than Diamond OA and comparable to Gold OA.

In Ukrainian journals, average FNCI values were substantially lower than in foreign journals, remaining below 0.4 (Figure 9). Nevertheless, highly cited outliers occurred mainly in Gold OA and Diamond OA.

**Access models, authorship patterns, and citation impact in foreign journals.**
Figure 10 shows the average FNCI of articles across access models and authorship patterns. It reveals a notable gap between internationally co-authored and nationally authored articles. For NASU, this gap was most pronounced in Bronze OA, Diamond OA, and Hybrid Gold OA, where internationally co-authored articles had an average FNCI around or above the world average.

A similar pattern is observed for universities, with internationally co-authored Bronze OA, Hybrid Gold OA, Gold OA, and Green OA articles reaching or exceeding the world average. However, these high averages were largely driven by highly cited outliers, as was also observed for NASU. The average FNCI of nationally authored articles was substantially lower than that of internationally co-authored articles, except for Bronze OA (0.89) in 2022–2023. However, both NASU and universities had highly cited outliers among domestically authored articles (Figure 11).

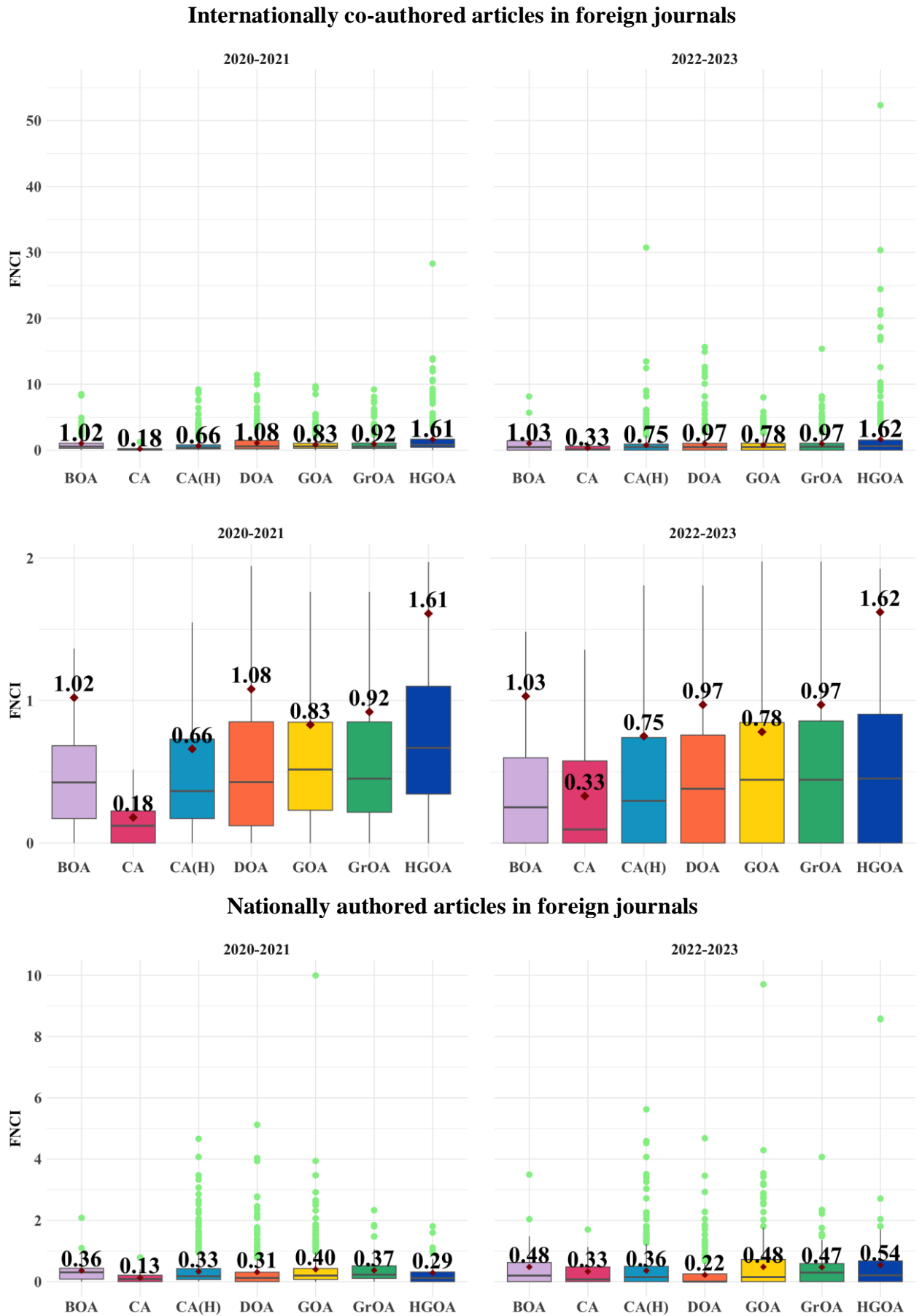

Fig. 10 Citation impact across authorship patterns and access models in foreign journals (NASU)

**Internationally co-authored articles in foreign journals**

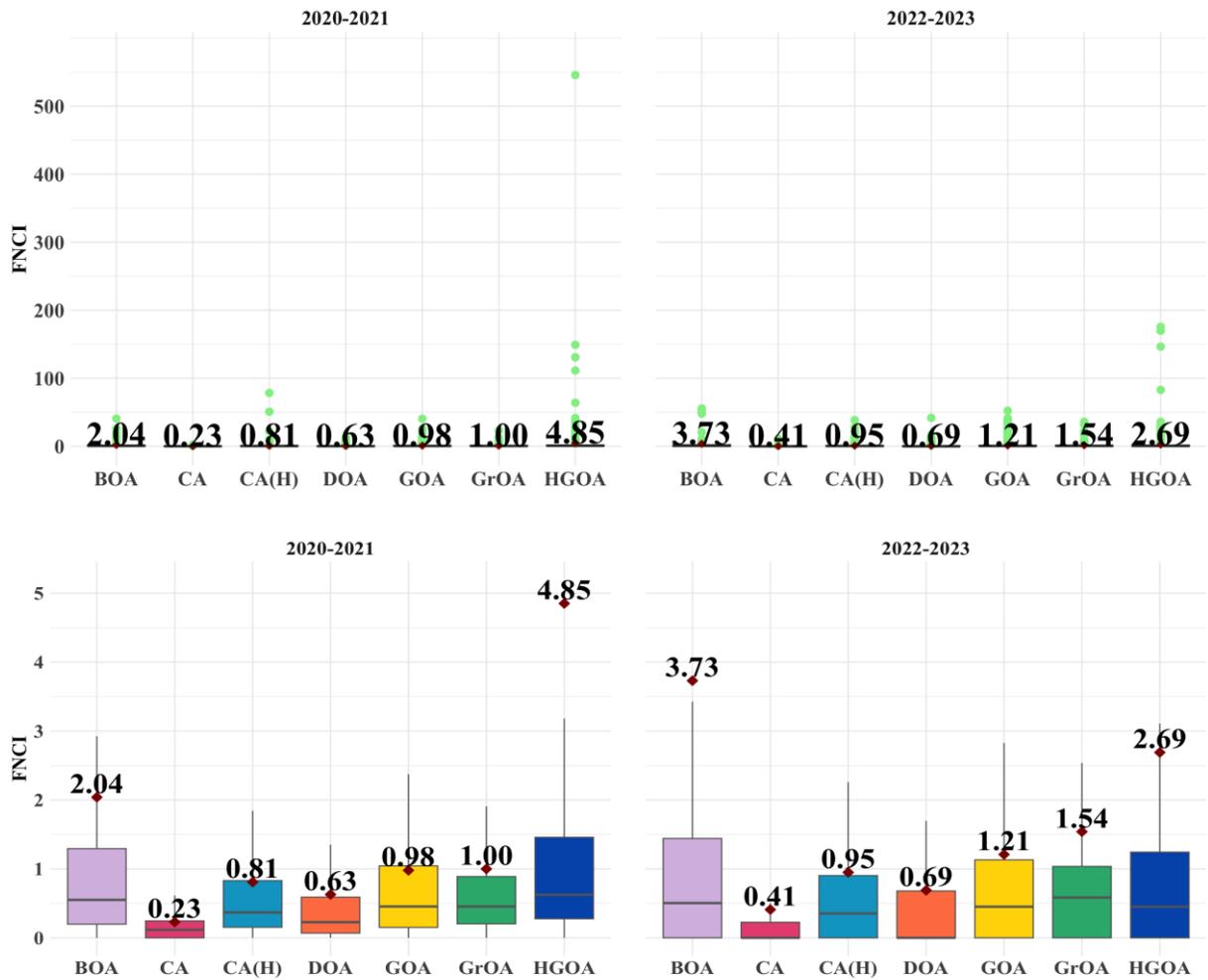

**Nationally authored articles in foreign journals**

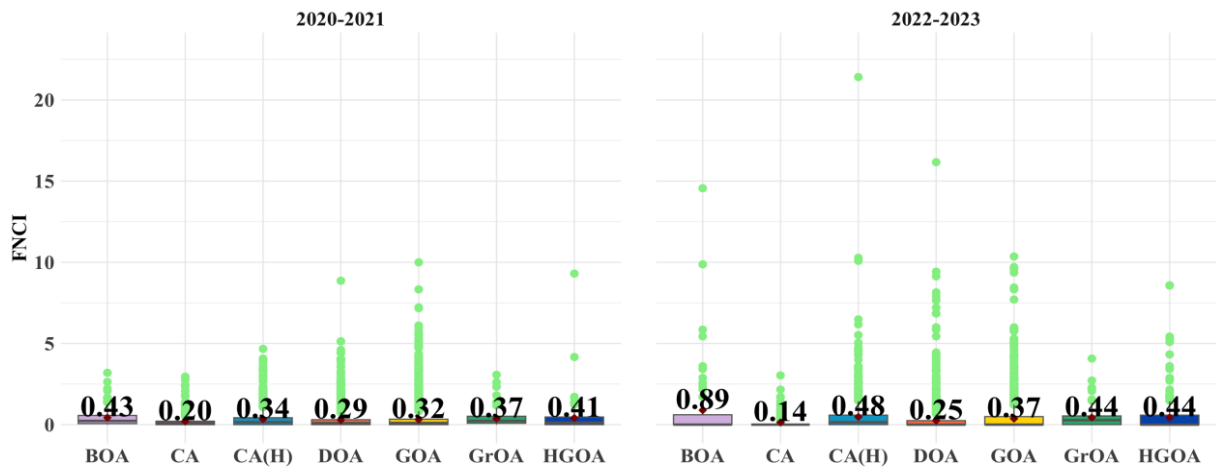

Fig. 11 Citation impact across authorship patterns and access models in foreign journals (universities)

published by universities in 2022-2023 is higher than the NASU. NASU outperforms universities in the citation impact of Diamond OA and Hybrid Gold OA, Gold OA and Green OA articles authored solely by scholars with Ukrainian affiliation.

Negative binomial regressions were estimated separately for NASU institutes (Model 1) and universities (Model 2), using Bronze OA publications with international co-authorship in international journals as the baseline category (Table 1). For NASU, the intercept was not significantly different from the reference group, indicating no difference in the baseline category. In contrast, for universities, the intercept IRR = 1.75 (95% CI: 1.55–1.98), meaning that university publications in the baseline category had 75% higher expected counts compared with the reference unit.

The time period 2022–2023 did not yield a significant effect for NASU, while for universities it was associated with a modest 13% increase. Access model showed strong and consistent effects: closed access and Diamond OA were linked to substantially lower counts, most notably –77% for closed access in universities. Hybrid Gold OA, by contrast, exhibited strong positive associations in both models (+69% in NASU and +55% in universities). Ukrainian authorship and publication in Ukrainian journals were consistently associated with significantly lower expected counts (reductions of ~50% or more). The models explained ~21% of deviance overall, with stronger explanatory power in universities due to the larger number of observations.

Table 1 Access model, authorship pattern, journal origin, and FNCI (negative binomial regression)

| Variable | (1) NASU | | | | (2) Universities | | | |
|---|---|---|---|---|---|---|---|---|
| | Estimate | IRR | Change (%) | 95% CI (IRR) | Estimate | IRR | Change(%) | 95% CI (IRR) |
| Intercept (baseline: Bronze OA, international authorship, international journal) | -0.097 (0.2285) | 0.91 | n.s. | 0.76-1.080 | 0.562*** (<2e-16) | 1.75 | +75% | 1.55-1.98 |
| 2022-2023 | 0.020 (0.431) | 1.02 | n.s. | 0.97-1.07 | 0.119*** (2.30e-11) | 1.13 | +13% | 1.09-1.17 |
| Closed access | -0.983*** (0.431) | 0.37 | -63% | 0.23-0.58 | -1.464*** (<2e-16) | 0.23 | -77% | 0.19-0.29 |
| Closed access in hybrid journal | -0.225* (0.015) | 0.80 | -20% | 0.67-0.96 | -0.674*** (<2e-16) | 0.51 | -49% | 0.45-0.58 |
| Diamond OA | -0.214* (0.022) | 0.81 | -19% | 0.68-0.97 | -1.004*** (<2e-16) | 0.37 | -63% | 0.32-0.42 |
| Gold OA | 0.026 (0.780) | 1.03 | n.s. | 0.86-1.23 | -0.568*** (<2e-16) | 0.57 | -43% | 0.50-0.64 |
| Green OA | 0.014 (0.890) | 1.01 | n.s. | 0.84-1.23 | -0.440*** (<2e-16) | 0.64 | -36% | 0.56-0.75 |
| Hybrid Gold OA | 0.523*** (4.65e-08) | 1.69 | +69% | 1.40-2.04 | 0.438*** (5.09e-10) | 1.55 | +55% | 1.36-1.77 |
| Ukrainian authorship | -0.779*** (<2e-16) | 0.46 | -54% | 0.43-0.49 | -0.100*** (1.20e-08) | 0.37 | -63% | 0.35-0.38 |
| Ukrainian journal | -1.001*** (<2e-16) | 0.37 | -63% | 0.34-0.40 | -0.713*** (<2e-16) | 0.49 | -51% | 0.47-0.51 |
| Deviance explained | 0.2103 | | | | 0.2072 | | | |
| Obs. | 17553 | | | | 45496 | | | |

**Distribution of the top10% most cited globally articles by access model and the journal origin**

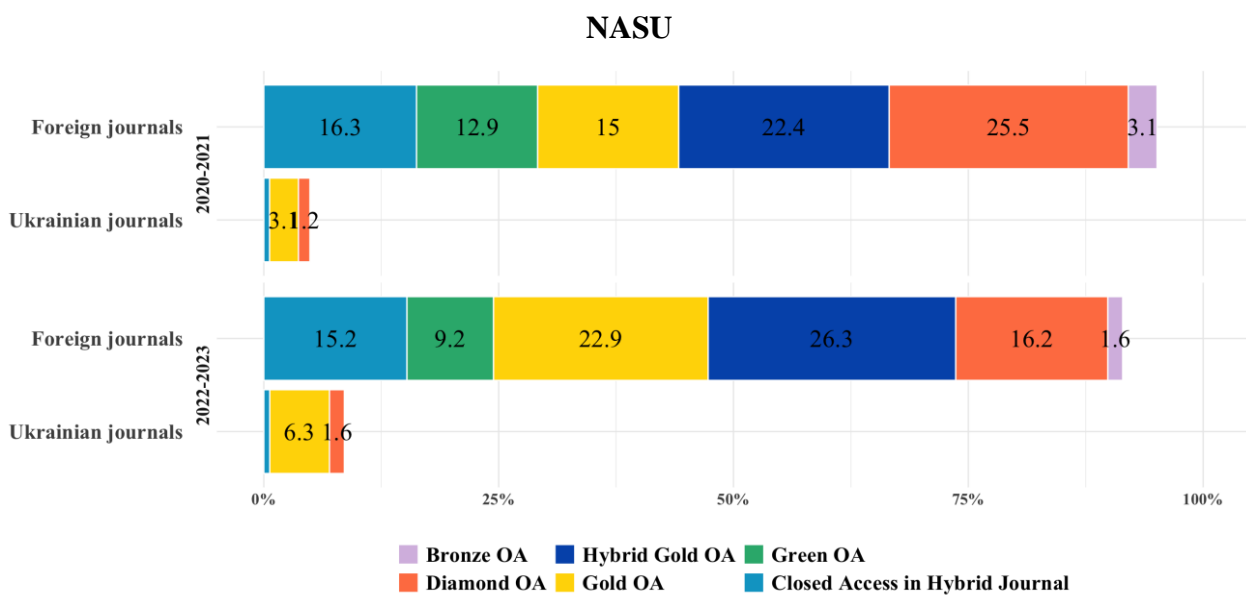

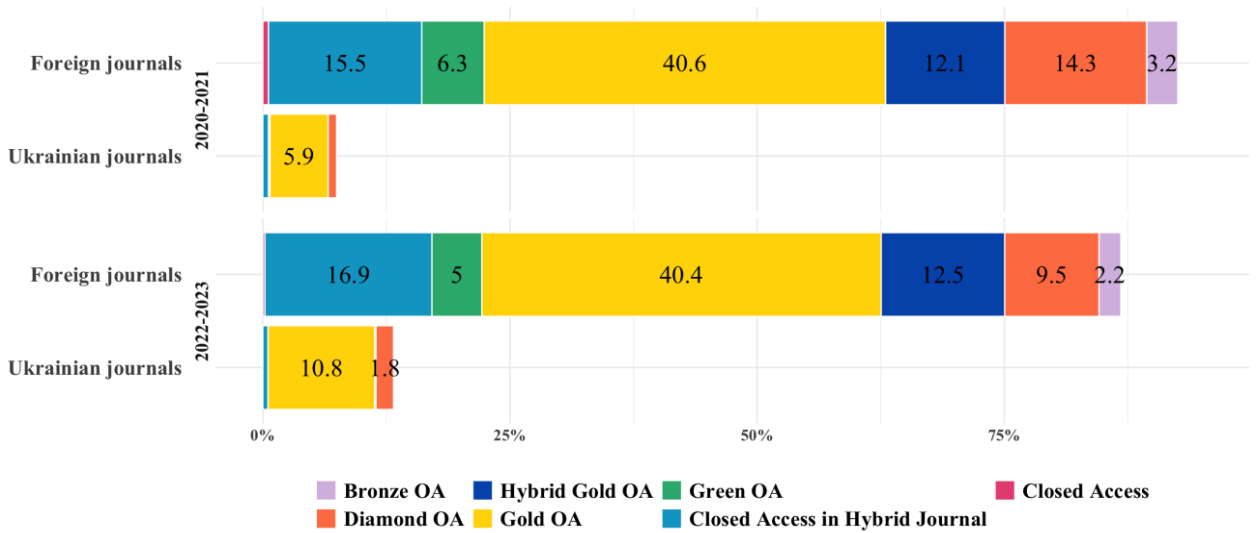

Fig. 12 Distribution of top 10% most cited globally by access model and journal origin

Figure 12 illustrates the distribution of the top 10% most cited globally articles by access model and journal origin. Articles in foreign journals constituted the majority of the top 10% most cited globally for both NASU and universities in both periods. The share of the top 10% most cited globally articles in Ukrainian journals was higher for universities than for NASU. In 2022-2023, it rose for both and reached 8.7% for NASU and 13.2% for universities. For NASU, the 10% most cited globally articles were rather balanced among access models in 2020-2021, while Gold OA (22.9%) and Hybrid Gold OA (26.3%) outperformed Diamond OA (16.3%). For universities, Gold OA articles accounted for the largest share of the top 10% most-cited articles globally in both periods, likely reflecting that Gold OA comprised the largest portion of university research output. In 2022–2023, the share of Gold OA articles among the top-cited publications declined slightly, from 43.0% to 41.1%. They were followed by closed-access articles in hybrid journals.

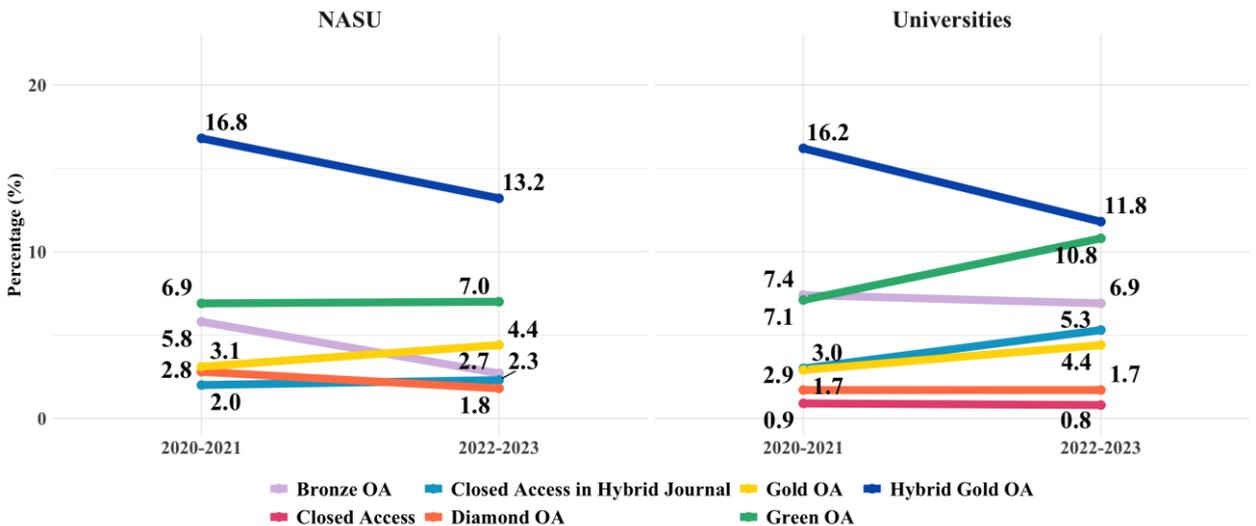

Fig. 13 The percentage of the top 10% most cited globally in each access model in foreign journals

Figure 13 illustrates the percentage of the top 10% most cited articles globally within each access model in foreign journals. For NASU, the highest percentage of top 10% most cited articles was observed in Hybrid Gold OA and Green OA. While the share in Hybrid Gold OA declined slightly over time, it increased in Gold OA. For universities, the percentage of top 10% most cited articles similarly decreased in Hybrid Gold OA but rose in Green OA, closed access in hybrid journals, and Gold OA.

## Conclusions

This study aimed to explore, first, the relationship between access model, authorship pattern, and citation impact using articles published by Ukrainian scholars between 2020 and 2023, and second, how Russia's full-scale war against Ukraine affected OA scholarly publishing. First and foremost, Ukrainian scholars deserve respect for continuing their research and publishing enduring life-threatening conditions, mental strain, shelling, blackouts and economic recession.

### *Access models and authorship patterns*

OA articles accounted for the majority of publications by both NASU and university-affiliated scholars, reaching 75.4% and 85.8%, respectively (83.8% overall). This share exceeds levels reported for Germany, the UK, and Brazil (Shu & Lavier, 2024). The growth was driven by Gold OA and Hybrid Gold**.** The results revealed that Diamond OA was the prevalent access model among NASU publications, despite physics and engineering being its main research field. This finding contrasts with prior studies, which have emphasised that Diamond OA is more prevalent in the social sciences (Bosman et al. 2021; Taubert et al. 2024; van Bellen and Céspedes 2024; Zhang et al. 2020). The share of Diamond OA in the total research output of NASU was relatively small, making Gold OA the predominant access model, accounting for 44.5% of publications in 2023.

In foreign journals, the percentage of internationally co-authored articles was much higher in Gold OA and Hybrid Gold OA than in Diamond OA, underscoring the importance of international collaboration for covering APCs. This aligns with Bordons et al. (2023) on the positive relationship between internationality of research and APC-based OA.

### *Implications of Russia's full-scale invasion of Ukraine*

The war contributed to (1) a decline in the share of articles in foreign journals for both NASU and universities, (2) a decrease in Gold OA in foreign journals with a simultaneous increase in Gold OA in Ukrainian journals for universities, and (3) a rise in internationally co-authored Gold OA articles in foreign journals for both entities. These trends reflect scholars' reduced ability to pay APCs due to the economic downturn.

While Elsevier's temporary waivers for Hybrid Gold OA led to an increased number of articles with corresponding authors affiliated with Ukrainian institutions, waivers for Gold OA provided by Elsevier and Springer also boosted publication, with the effect more pronounced for Elsevier. Nevertheless, MDPI and Aluna remained the top Gold OA publishers, likely due to reduced APCs provided by Aluna and vouchers covering APCs for MDPI, rapid publication timelines, and low rejection rates. Limited transparency around waiver policies provided by major publishers and the risk of mid-review changes further added uncertainty to an already precarious academic environment in Ukraine.

### *Citation impact across access models*

Bronze OA, Hybrid Gold OA, and Green OA articles published in hybrid journals were the most cited for both universities and NASU, with these access models also accounting for the highest share of the top 10% most cited articles globally. These findings align with previous studies showing that Green and Hybrid Gold OA are among the most cited OA models (Piwowar, 2018; Dorta-González & Dorta-González, 2022), likely because hybrid journals are well-established and widely recognised (Science-Metrix, 2018; Demeter et al., 2021). Taking into account the citation advantages of publishing in hybrid journals, Green OA through preprint posting is considered the most sustainable OA route prioritised by both EC policies. The finding also aligns with prior studies revealing that Bronze OA, Hybrid Gold OA and Green OA are related to higher citation advantages than Gold OA (Pinfield et al., 2017; Björk & Solomon, 2017; Brainard, 2024; Maddi & Sapinho, 2021).

Green OA, Bronze OA, and Hybrid Gold OA generally received higher citation counts than closed-access articles. However, Diamond OA articles were cited less than closed-access articles in hybrid journals for universities, which may partly reflect incomplete Scopus data on preprints. As noted by Gargouri et al. (2010), OA tends to increase citation impact primarily for articles that already have inherent citation potential. The lower citation impact of Diamond OA compared to Gold OA suggests the need for further investigation into the factors limiting its visibility.

Statistical models confirmed a positive correlation between Bronze OA, Hybrid Gold OA, international co-authorship, publishing in foreign journals, and citation impact, though they explained only about 20% of the variance, consistent with prior findings that citation impact depends on multiple factors, including topic novelty (Tahamtan et al., 2016), and funding (Dorta-González & Dorta-González, 2022). Despite this, high-impact outliers were observed among both internationally co-authored and nationally authored articles published in both foreign and Ukrainian journals.

*Ukrainian journals*
In 2022–2023, the share of articles published in Ukrainian journals was approximately 54.9% for NASU and 40% for universities, with 85.9% of NASU articles and 87.5% of university articles authored solely by national scholars. The average FNCI of articles in Ukrainian journals remained lower than that of foreign journals. However, the number of highly cited outliers increased, and the share of Ukrainian journal articles among the top 10% most cited globally also grew; for universities, this share reached 13.2%. These findings suggest that Ukrainian journals should focus on strengthening peer review to enhance article quality, thereby increasing attractiveness for international submissions, and actively promote their journals abroad to foster international collaboration, which can further improve the quality and impact of published research.

**Concluding reflections and policy directions**
The study confirms the growing dominance of Gold OA, consistent with prior research showing that an increasing number of journals are transitioning to this model and that Gold OA quantitatively prevails over other OA models (Butler et al., 2023; Warren, 2025; Wiley, 2025; Springer, 2019; Levin et al., 2024). However, this trend also poses significant risks and negative consequences for global science (Singh et al., 2021; Levin et al., 2024). Publisher fee waivers, while valuable in the short term, remain uncertain and cannot provide a reliable foundation for long-term research planning. The fact that Ukrainian scholars under wartime conditions continue to prioritise outlets such as MDPI and Aluna, allocating already diminished financial resources to APCs, raises further concerns.

As APC-based OA does not represent a sustainable use of research funding (Haustein et al., 2024; Alonso-Álvarez et al., 2024), the European Commission (2017), through its OpenAIRE infrastructure, as well as advanced economies such as Belgium (FWO, 2021), Italy (Picarra et al., 2015), and Japan (Chawla, 2024), have emphasised the importance of Green OA over Gold OA. These examples demonstrate that sustainable approaches are feasible, strengthening access to research while minimising financial barriers for scholars.

The finding that Diamond OA publications achieve lower citation impact than Gold OA has policy implications. It underscores the need to make Diamond OA not only equitable but also competitive in terms of visibility and impact. Policy directions should therefore include targeted support for Diamond journals in areas such as indexing in major databases, robust quality assurance mechanisms, and sustainable funding models.

To summarise, scholarly publishing requires OA sustainable models that ensure equity, rapid publication timelines and broad dissemination of research.

**Acknowledgements**
This project has received funding through the MSCA4Ukraine project, which is funded by the European Union. Views and opinions expressed are however those of the author only and do not necessarily reflect those of the European Union. Neither the European Union nor the MSCA4Ukraine Consortium as a whole nor any individual member institutions of the MSCA4Ukraine Consortium can be held responsible for them.

Kiselyova and Ivashchenko 2024; Oleksiyenko et al 2021; 2023; Lutsenko et al 2025; Fiialka, 2022

Nazarovets and Mryglod 2025;

Hladchenko and Westerheijden 2019; Hladchenko 2023; Hladchenko 2024